\documentstyle[12pt,preprint]{aastex}

\begin{document}

\title{Fixed-delay Interferometry for Doppler Extra-solar Planet Detection}

\author{Jian Ge}

\affil{Department of Astronomy and Astrophysics,  Penn State
University, University Park, PA 16802; Email: jian@astro.psu.edu}

\begin{abstract}

We present a new technique based on fixed-delay interferometry for
high throughput, high precision and multi-object Doppler radial
velocity (RV) surveys for extra-solar planets. The Doppler
measurements are conducted through monitoring the stellar fringe
phase shifts of the interferometer instead of absorption line
centroid shifts as in state-of-the-art echelle spectroscopy. High
Doppler sensitivity is achieved through optimizing the optical
delay in the interferometer and reducing photon noise by measuring
multiple fringes over a broadband. This broadband operation is
performed through coupling the interferometer with a low to medium
resolution post-disperser. The resulting fringing spectra over the
band pass are recorded on a 2-D detector, with fringes sampled in
the slit spatial direction and the spectrum sampled in the
dispersion direction. The resulting total Doppler sensitivity is,
in theory, independent of dispersing power of the post-disperser,
which allows development of new generation RV machines with much
reduced size, high stability and low cost compared to echelles.
This technique has the potential to improve RV survey efficiency
by 2-3 orders of magnitude over cross-dispersed echelle
spectroscopy approach to allow a full sky RV survey of hundreds of
thousands of stars for planets, brown dwarfs, and stellar
companions once the instrument is operated as a multi-object
instrument and optimized for high throughput. The simple
interferometer response potentially allows this technique to be
operated at other wavelengths independent of popular iodine
reference sources, being actively used in most of the current
echelles for Doppler planet searches, to search for planets around
early type stars, white dwarfs, and M, L and T dwarfs for the
first time. The high throughput of this instrument could also
allow investigation of extra-galactic objects for RV variations at
high precision.

\end{abstract}



\section{Introduction}

Interferometer techniques based on variable optical delays have
already been proposed for high precision RV measurements (Connes
1985; Fradsen et al. 1993; Douglas 1997), the most recent of which
is Holographic Heterodyning Spectroscopy (HHS). To our knowledge,
none of these techniques has yet achieved RV precision on a par
with cross-dispersed echelle spectroscopy ($\sim$ 3 m/s)(Butler et
al. 1996). The fundamental limitation of interferometric
techniques is the narrow band pass (e.g. $\sim$ 30 \AA) compared
to the broadband operation of the echelle ($\sim$ 1000 \AA),
which provides $\sim$ 6 times higher Doppler error than the
echelles at the same spectral resolution.

The narrow band limitation can be overcome by a new kind of
interferometer approach based on a fixed optical delay and a
post-disperser. Fixed-delay interferometers with narrow band
passes have been used in high precision RV measurements in solar
astrophysics since 1980's (Title \& Ramsey 1980; Harvey et al.
1995; Kozhevatov e al. 1995; 1996). The best RV precision of
$\sim$ 3 m/s has been reported with solar observations (Kozhevatov
e al. 1995; 1996). Recent laboratory work with a wide angle
Michelson interferometer with a fixed delay and a medium
resolution grating post-disperser appears to offer $\sim$ 1 m/s RV
precision, similar to the echelle spectroscopy (Erskine \& Ge
2000). The first light of a prototype instrument based on this
concept at the Hobby-Eberly 9 m and Palomar 5 m telescopes in 2001
demonstrates that $\sim$ 8 m/s RV precision 
has been achieved with stellar observations (Ge et
al. 2001; van Eyken et al. 2001; Ge et al. 2002). Here we present
the theoretical principle behind this new RV technique, its
performance comparison with the cross-dispersed echelle technique
and its new capability for all sky RV surveys for extra-solar
planets.

\section{Principle of Fixed-delay Interferometry}

This new approach is illustrated in Figure 1. The circular
incoming beam from a telescope is converted to a rectangular one
by cylindrical optics, split into two beams with equal amplitude
and fed to an interferometer with a fixed optical delay in one of
the arms to form fringes in stellar absorption lines. The
uncorrelated fringes over a broadband (white fringes) are
separated by a post-disperser and recorded on a 2-D detector array
with fringes sampled on the slit spatial direction and the
spectrum sampled in the dispersion direction. This post-disperser
can be any dispersing device (e.g. grating, prism, grism,
Fabry-Perot interferometer). The fringe period in the slit spatial
direction is proportional to wavelength. Therefore, fringes over a
bandwidth  similar to, or larger than, that of the echelles can be
properly sampled and measured on a suitable detector array
(e.g. CCD), which can help reduce RV measurement
noise due to photon statistics.

In this interferometer, high contrast interference fringes are
formed when the optical path difference  is within the coherence
length (Goodman 1984). Parallel fringes are created when the faces
of the two interferometer mirrors are slightly tilted with respect
to each other. The interference order, $m$, is determined by
\begin{equation}
m\lambda =d,
\end{equation} where $d$ is the total optical delay, and $\lambda$
is the operating wavelength. Once the delay is fixed, Doppler RV
motion will shift the fringes to different orders. The
corresponding Doppler velocity shift, $\Delta v$, is
\begin{equation}
\Delta v=\frac{c\lambda}{d}\Delta m.
\end{equation}
The Doppler analysis is performed on the$``$normalized" fringe
(i.e., divided by the continuum), which can be represented as a
function of velocity
\begin{equation}
I=1+\gamma_i sin(2\pi\frac{v}{c\lambda/d}+\phi_0),
\end{equation}
 where $\gamma_i$  is the fringe visibility, and $\phi_0$ is the phase value for the
  first pixel of the fringe. Along the fringe, the average
  uncertainty in the velocity from one pixel is
\begin{equation}
\sigma_p=<\frac{\epsilon_I}{dI/dv}>\approx\frac{c\lambda}{4d\gamma_i(S/N)},
\end{equation}
 where $\epsilon_I$ is the uncertainty in the residual
intensity at the pixel, for a normalized spectrum, $\epsilon_I
=1/(S/N)$, where $(S/N)$ is the signal-to-noise ratio at the
pixel, and the average slope over one side of the fringe (half
period) of the sinusoidal interferometer response
\begin{equation}
<\frac{dI}{dv}>=\frac{4d\gamma_i}{c\lambda}
\end{equation}
 is used.
If an interference fringe is sampled by $N_{pix}$ pixels, then the
total intrinsic Doppler error under photon noise limits is
\begin{equation}
\sigma_{f,i}\approx\frac{c\lambda}{4d\gamma_i\sqrt{F_i}},
\end{equation}
 where $(S/N) =\sqrt{N_{ph}}$  is applied, $N_{ph}$ is the number of
photons received by each pixel, and $F_i = N_{pix}N_{ph}$ is the
total photon number collected by each fringe. It is clear that the
Doppler precision in the interferometer approach is determined by
delay, visibility, and total photons collected in each fringe. If
we simply assume that each stellar intrinsic absorption line is a
Gaussian shape with a FWHM of $\Delta \lambda_i$ and a depth of
$D_i$ ($0\leq D_i\leq 1$), then the visibility as a function of
optical delay can be derived (Goodman et al. 1984) as
\begin{equation}
\gamma_i=D_i e^{-3.56\frac{d^2}{l_c^2}},
\end{equation}
 where $l_c = \lambda^2/\Delta \lambda_i$ is the coherence length of
 the interferometer beam. A
simple derivation shows that $(d\gamma_i)$ reaches a maximum value
of $0.23D_il_c$, when $d = 0.37l_c$ as shown in Figure 2. Therefore, the minimum
intrinsic Doppler error per fringe is
\begin{equation}
\sigma_{f,i}\approx\frac{1.1c\lambda}{D_il_c\sqrt{F_i}},
\end{equation}
This formula indicates that the intrinsic Doppler error for each
fringe decreases with increasing coherence length of the light,
the flux per fringe and also with increasing absorption line
depth. If multiple fringes ($N_i$) over a broadband are used for
Doppler RV measurements, then the total Doppler error is
\begin{equation}
\sigma_{f,i,t}\approx\frac{1}{\sqrt{N_i}}\frac{1.1c\lambda}{D_il_c\sqrt{F_i}},
\end{equation}
assuming line depths, widths, and $F_i$ are the same for all
fringes. For the multiple fringe measurements, the intrinsic line
profiles are convolved with the response of the post-disperser,
which affects RV precision. Assuming that the post-disperser
response profile is approximately a Gaussian function with a FWHM
of $\Delta \lambda_g$, the FWHM of the observed line is $\Delta
\lambda_o \approx \Delta \lambda_g$ if $\Delta \lambda_g >>\Delta
\lambda_i$. The observed absorption line depth is
\begin{equation}
D_o\approx \frac{\Delta\lambda_i}{\Delta\lambda_o}D_i.
\end{equation}
 The observed flux within each fringe (or absorption
line) is increased to
\begin{equation}
F_o\approx \frac{\Delta\lambda_o}{\Delta\lambda_i}F_i.
\end{equation}
  The total measured Doppler error per
fringe becomes
\begin{equation}
\sigma_{f,o}\approx
\sqrt{\frac{\Delta\lambda_o}{\Delta\lambda_i}}\sigma_{f,i}.
\end{equation}
The Doppler sensitivity for each fringe is decreased due to the
use of the post-disperser. However, if the detector dimension in
the dispersion direction and the sampling of each resolution
element are fixed, then the total number of absorption lines
covered by the array, $N_o$, increases to
\begin{equation}
N_o\approx \frac{\Delta\lambda_o}{\Delta\lambda_i}N_i,
\end{equation}
  if the absorption line
density is approximately constant over the wavelength coverage.
Therefore, the total Doppler error
\begin{equation}
\sigma_{f,o,t}\approx \frac{1}{\sqrt{N_o}}\frac{1.1c\lambda}{D_ol_c\sqrt{F_o}}\approx \sigma_{f,i,t}.
\end{equation}
is independent of the resolution of the post-disperser, which is
significantly different from the echelle approach and offers new
possibilities for Doppler planet searches.

In the cross-dispersed echelle spectroscopy, a total measured
Doppler error (photon noise error) for a stellar absorption line
with an intrinsic FWHM, $\Delta \lambda_i$, and depth, $D_i$, at a
spectral resolution of $\Delta \lambda_o$ is described as
\begin{equation}
\sigma_{e,o}\approx(\frac{\Delta\lambda_o}{\Delta\lambda_i})^{3/2}\frac{c\Delta
\lambda_i}{D_i\lambda\sqrt{F_i}}=
(\frac{\Delta\lambda_o}{\Delta\lambda_i})^{3/2}\sigma_{e,i},
\end{equation}
where
$\Delta\lambda_o=\sqrt{\Delta\lambda_e^2+\Delta\lambda_i^2}$,
$\Delta \lambda_e$ is the FWMH of the echelle response.
For a fixed bandwidth  of the echelle, e.g.,
$\sim$ 1000
\AA\ determined by the bandwidth of iodine absorption in the Visible for calibrations
 (Butler et al. 1996), the total measured echelle error is
\begin{equation}
\sigma_{e,o,t}\approx(\frac{\Delta\lambda_o}{\Delta\lambda_i})^{3/2}\sqrt{\frac{N_i}{N_e}}\sigma_{e,i,t}
\approx (\frac{\Delta\lambda_o}{\Delta\lambda_i})^{3/2}\sqrt{\frac{N_i}{N_e}}\sigma_{f,o,t},
\end{equation}
where the total number of stellar lines covered in this band,
$N_e$, is fixed. This indicates that the Doppler error in the
echelle approach strongly depends on the echelle resolving power.
At typical echelle resolution such as $R \sim$ 60,000 (e.g.
Suntzeff et al. 1994; Vogt et al. 1994; D'Odorico et al. 2000),
the Doppler error is $\sim$ 1.3 times higher than that in the
interferometer approach for the same wavelength coverage and
photon flux. Therefore, in order to approach the highest possible
Doppler precision, approximately the same as that in the
interferometer approach,  limited only by the stellar intrinsic
line profiles, the spectral resolving power must be much higher
than the stellar intrinsic line width, i.e.
$\Delta\lambda_e<<\Delta\lambda_i$.

\section{Discussions}

The independence of Doppler sensitivity with the post-disperser
resolving power in the interferometer approach potentially allows
about two to three orders of magnitude improvement in RV survey speed
for  planets over echelle techniques. The use of low resolution but
 high efficiency post-dispersers can significantly boost the overall
 detection efficiency and allow single dispersion order operations for
 potential multiple object observations. The current echelle instruments
 are limited to a few percent total detection efficiency, or less; this
 includes telescope transmission, slit loss, spectrograph and detection
 loss, due to use of very high resolution echelle and complicated camera
 optics (Vogt et al. 1994; Suntzeff et al 1996; D'Odorico et al. 2000).
 An interferometer coupled with a low resolution dispersing
 instrument potentially offers perhaps $\sim$ 30\%
 detection efficiency, 5-10 times higher than the echelle, thus
 allowing to extend RV survey depth to fainter objects with fixed
 size telescope. This high efficiency can be achieved
 because both of the interferometer and low resolution
 spectrograph can be optimized for high transmission\footnote{For
 instance, a Michelson type interferometer with corner cube
 mirrors can feed both interferometer outputs to the detector at
 $\sim$ 90\% efficiency (Traub 2002). A low resolution spectrograph using a
 volume phase grating can potentially reach $\sim$ 70\%
 transmission (Barden et al. 2000). Together with the telescope
 transmission ($\sim$ 80\%), fiber-feed transmission ($\sim$ 70\%)
 and detector quantum efficiency ($\sim$ 90\%), the total
 detection efficiency can reach $\sim$ 30\%. Details about the total
 transmission budget will be discussed in a follow-up paper (Ge et
 al, 2002)}. The potential operation of the instrument in a
 multi-object mode allows simultaneous observations of hundreds
 of objects in a single exposure with broadband coverage on a large
  2-D detector array.  Full sky coverage of an RV survey for planets
  becomes possible with a wide field telescope.

In addition, the simple response function potentially offers lower
systematic errors than those echelle approaches. Currently, the
systematic errors associated with the echelle instrument response
account for about 2 m/s Doppler error largely due to the
de-convolution of observed stellar spectra to create star
templates (Butler et al. 1996; Valenti 2000). Since this process is
not required in this interferometer approach, the systematic error
can be well below 2 m/s. Hence, Doppler precision of sub m/s is
potentially reachable through increasing photons collected by each
fringe and increasing wavelength coverage.  We have achieved $\sim$ 8 m/s
Doppler precision with star light at the HET with an approximately $\sim$ 140 
\AA\ wavelength coverage, a $R = 6700$ post-disperser and a S/N 
$\sim$ 100 per pixel.  This error is consistent with
photon-noise limit\footnote{The reason that we were not able to 
reach higher precision is that the calibration error from using iodine 
reference is $\sim$ 7 m/s, dominating the total measurement error. At
$R \sim$ 6700, mean fringe visibility for iodine absorption lines are
too low ($\sim$ 2.5\%) compared to that for stellar absorption lines ($\sim$
7\%). In the future, a calibration source with much higher 
fringe visibility will replace the iodine for achieving sub m/s. Details on 
new calibration techniques will be reported in the follow-up paper 
(Ge et al. 2002)}. 

Another exciting possibility with this interferometer technique is
to extend RV surveys for planets in wavelengths other than the
visible, previously not covered by echelle surveys.  Since the
interferometer response is simple and stable, there is no need for
calibrating instrument response as for the echelle. Instrument
wavelength calibrations (or instrument zero velocity drift
measurements) can be conducted with reference sources with a lower
line density than the iodine used in the echelle. Therefore, this
instrument can be easily adapted to other wavelengths for
maximizing the photon flux, and number of stellar absorption lines
for precision Doppler RV measurements. The candidate stars for
this potential survey include late M, L, and T dwarfs, early type
B, A stars and white dwarfs. Late M, L and T dwarfs have peak
fluxes in the near-IR. For instance, the late M dwarfs have peak
flux in the near-IR, at least a factor 10 higher than at the
visible (Kirkpatrick et al. 1993; 1999) and since a number of
molecular absorption lines are concentrated in this wavelength
region, observing time can be significantly reduced if the IR
spectra can be used for RV measurements. A and B main sequence
stars have very broad intrinsic absorption lines dominated by the
Balmer series due to rapid rotation.  White dwarfs have very broad
intrinsic absorption lines due to pressure broadening. Typical
rotation velocity for normal A and B type stars is about 150 km/s,
$\sim$ 30 times faster than a typical late type star (Gray 1992;
Dravins 1987). Due to the broader intrinsic line profile, the
intrinsic Doppler error for these stars increases by $\sim$ 6
times compared to the late type stars. However, with deep Balmer
absorption lines over large
wavelength coverage of most of the Balmer lines and high
signal-to-noise ratio data, it is possible to achieve $\sim$ a few
m/s Doppler precision for detecting planets around these stars for
the first time.

\acknowledgments

The author is grateful to David Erskine, Julian van Eyken, Suvrath
Mahadevan, Larry Ramsey, Don Schneider, Steinn Sigurdsson,  Web
Traub,  Ron Reynolds, Fred Roesler, Stuart Shaklan, Harvey
Moseley, Bruce Woodgate, Roger Angel, Mike Shao, Chas Beichman, Ed
Jenkins \& Jim Gunn for stimulating discussions on this new
instrument concept.

\newpage

\centerline {\bf Figure Captions}

\noindent Figure 1. -- Principle of a fixed-delay interferometer
and a post-disperser. The fringe data was taken with a prototype
instrument, called Exoplanet Tracker (ET) and a 1kx1k CCD array,
developed at Penn State(Ge et al. 2001; van Eyken et al. 2001).
 The stellar interference fringes,
formed by a Michelson type interferometer with $\sim$ 7 mm optical
delay, lie in the horizontal direction and are sampled by $\sim$
60 pixels. The fringes over different wavelengths are separated by
a first order diffraction grating with a resolving power of R
$\sim$ 6,000, a factor of ten times lower than typical echelle
spectrographs and recorded in the vertical direction of the CCD.

\noindent Figure 2. -- Product of delay and fringe visibility vs.
delay in the interferometer for a gaussian shaped profile.
The maximum for $d\gamma = 0.23D_il_c$ is at $d=0.37l_c$.
The optimal value for delay depends on shape of the stellar line profile.

\newpage

\begin{figure}
\vspace{4in}
\plotone{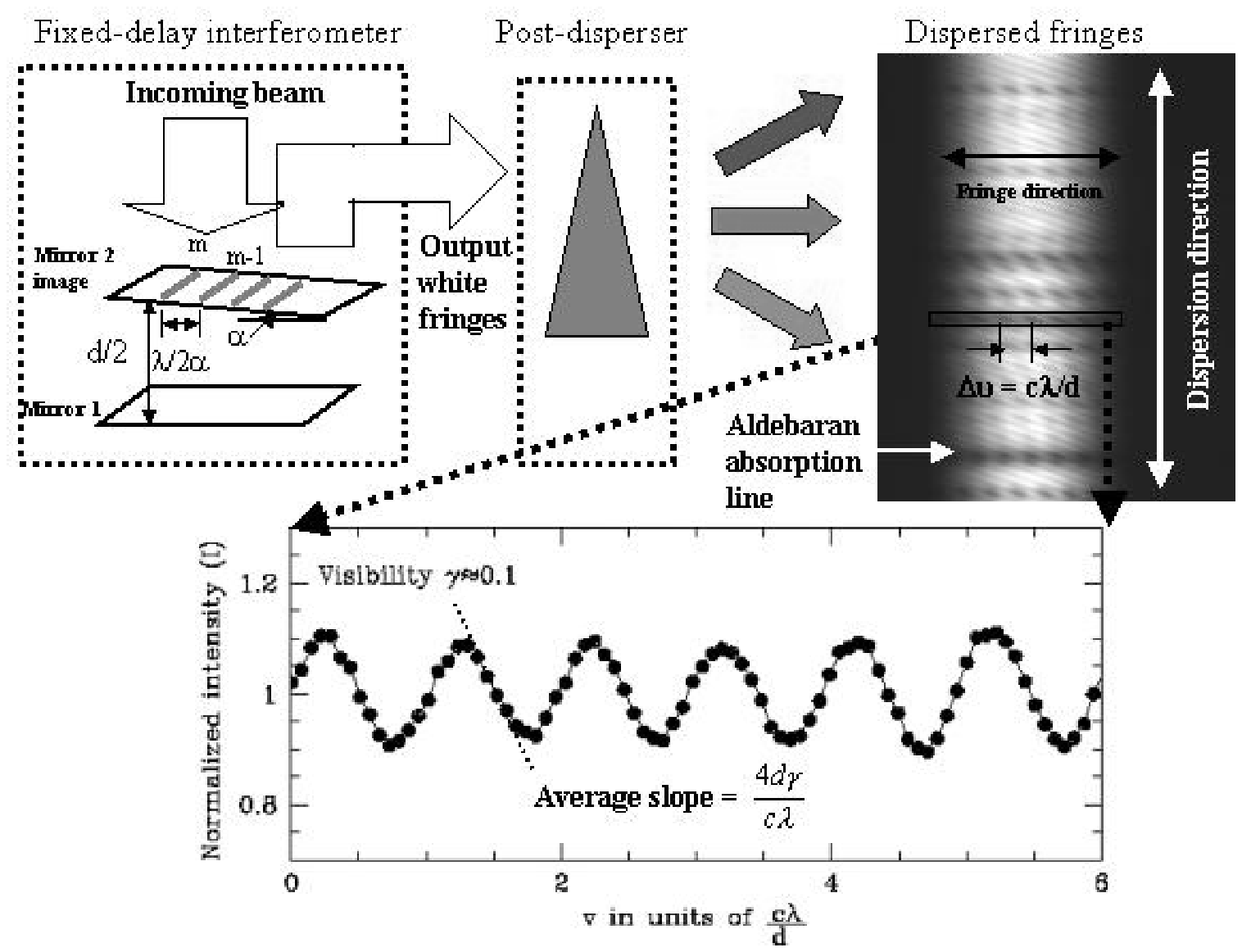}
\caption{}
\label{fig:fig1}
\end{figure}

\newpage
\begin{figure}
\plotone{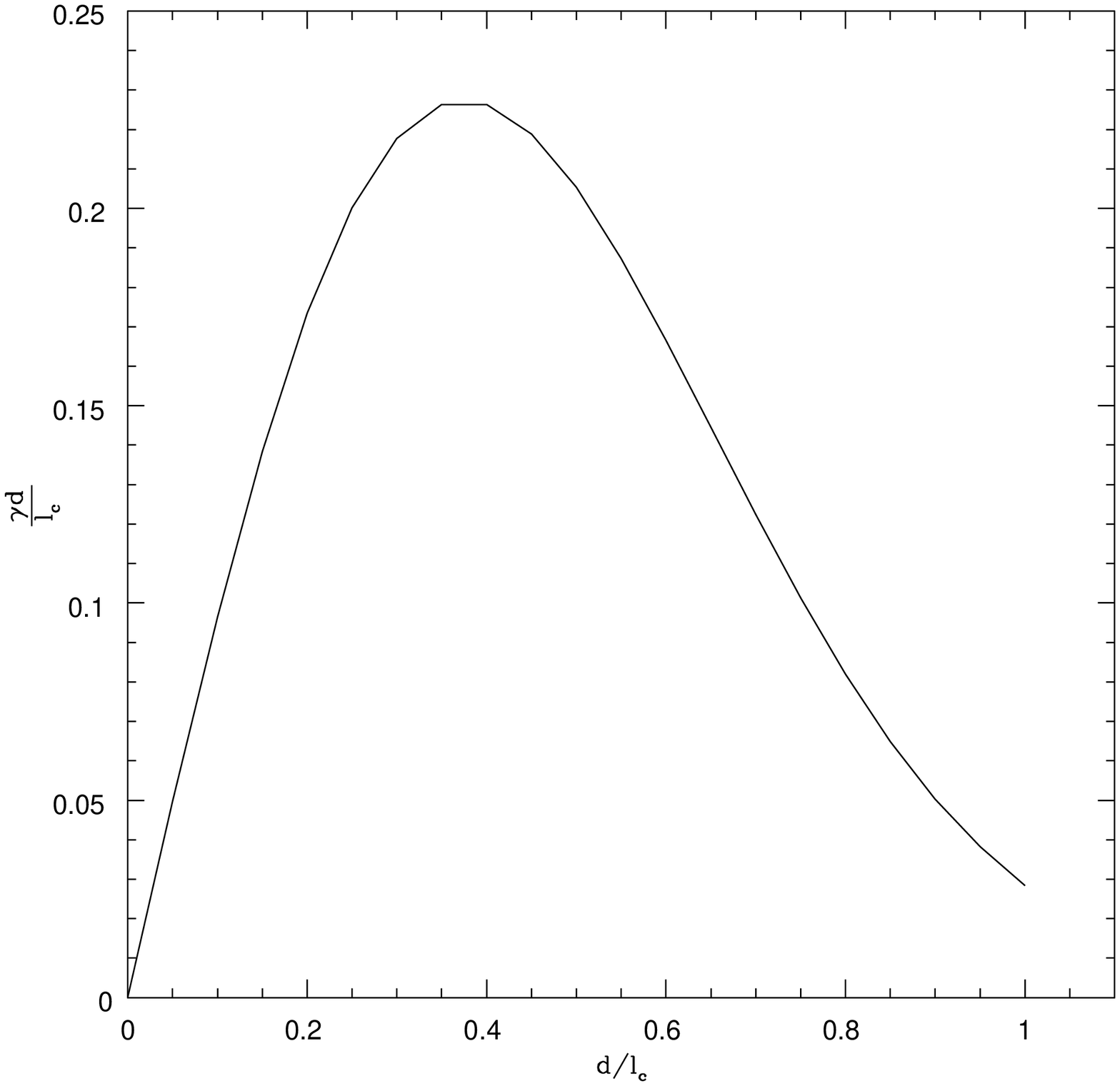}
\caption{}
\label{fig:fig2}
\end{figure}

\end{document}